# Privacy-Preserving Data in IoT-based Cloud Systems: A Comprehensive Survey with AI Integration

D. Dhinakaran[1], S.M. Udhaya Sankar[2], D. Selvaraj[3], S. Edwin Raja[4]

[1,4]Department of Computer Science and Engineering, Vel Tech Rangarajan Dr. Sagunthala R&D Institute of Science and Technology, Chennai, India

[2]Department of CSE (Cyber Security), R.M.K College of Engineering and Technology, Chennai, India

[3]Department of Electronics and Communication Engineering, Panimalar Engineering College, Chennai, India

[1]drdhinakarand@veltech.edu.in, [2]udhaya3@gmail.com, [3]mails2selvaraj@yahoo.com, [4]edwinrajas@gmail.com

*Corresponding author: D. Dhinakaran

**Abstract:**

As the integration of Internet of Things (IoT) devices with cloud computing proliferates, the paramount importance of privacy preservation comes to the forefront. This survey paper meticulously explores the landscape of privacy issues in the dynamic intersection of IoT and cloud systems. The comprehensive literature review synthesizes existing research, illuminating key challenges and discerning emerging trends in privacy-preserving techniques. The categorization of diverse approaches unveils a nuanced understanding of encryption techniques, anonymization strategies, access control mechanisms, and the burgeoning integration of artificial intelligence. Notable trends include the infusion of machine learning for dynamic anonymization, homomorphic encryption for secure computation, and AI-driven access control systems. The culmination of this survey contributes a holistic view, laying the groundwork for understanding the multifaceted strategies employed in securing sensitive data within IoT-based cloud environments. The insights garnered from this survey provide a valuable resource for researchers, practitioners, and policymakers navigating the complex terrain of privacy preservation in the evolving landscape of IoT and cloud computing.

**Keywords:** Internet of Things (IoT), cloud computing, privacy, encryption techniques, anonymization, access control.

## 1. Introduction

In recent years, the convergence of the Internet of Things (IoT) and cloud computing has ushered in a new era of connectivity and data-driven applications. The amalgamation of these two technological paradigms has not only opened avenues for innovation and efficiency but has also raised significant concerns about the privacy and security of the massive volumes of data generated and processed in these interconnected ecosystems [1]. This introduction aims to elucidate the pivotal role of privacy in the context of IoT-based cloud systems, provide a foundational understanding of IoT and cloud computing, shed light on the escalating concerns surrounding data privacy and security, and articulate the overarching objectives of this survey paper. The proliferation of IoT devices has permeated diverse aspects of our lives, from smart homes and wearable devices to industrial applications and smart cities [2]. These interconnected devices generate an unprecedented amount of data, ranging from personal information to critical industrial telemetry. Simultaneously, cloud computing has emerged as the backbone infrastructure supporting the storage, processing, and analysis of this deluge of data. However, as the volume and diversity of data grow, so does the importance of ensuring robust privacy measures [3].

Privacy in the context of IoT-based cloud systems is not merely a regulatory requirement but a fundamental ethical consideration. Users entrust a myriad of sensitive information to IoT devices, from health data collected by wearables to smart home configurations that reflect personal habits. This data, when transmitted to the cloud for storage and analysis, becomes part of a broader network where its security and privacy are paramount [4]. Unauthorized access or breaches can lead to severe consequences, ranging from identity theft and unauthorized surveillance to potential manipulation of critical infrastructure. To comprehend the significance of privacy in IoT-based cloud systems, it is essential to delve into the foundational concepts of IoT and cloud computing.

The IoT refers to the interconnection of physical devices, vehicles, buildings, and other items embedded with sensors, software, and network connectivity. These connected devices collect and exchange data, providing real-time insights and enabling intelligent decision-making. In the context of privacy, IoT devices act as data collectors that often handle sensitive information, necessitating robust privacy-preserving mechanisms. Cloud computing involves the delivery of computing services—such as storage, processing, and networking—over the internet [5]. It provides scalable and on-demand resources, allowing organizations to offload computational tasks and store large datasets efficiently. In the context of privacy, cloud computing introduces challenges related to data residency, access controls, and the security of data in transit and at rest.

## 1.1 Concerns about Data Privacy and Security

As the adoption of IoT-based cloud systems accelerates, concerns about data privacy and security have become increasingly pronounced. Several factors contribute to these apprehensions.

***Proliferation of Data:***

The sheer volume of data generated by IoT devices is staggering. From individual users' personal information to the telemetry data from industrial sensors, the scope and scale of data collection raise concerns about the potential misuse or unauthorized access.

***Data Accessibility:***

The accessibility of data in the cloud poses challenges related to who can access, manipulate, and analyze the information. Unauthorized access or data breaches can compromise the privacy of individuals and organizations alike.

***Interconnected Ecosystems:***

The interconnected nature of IoT-based cloud systems introduces complex relationships between devices, applications, and cloud infrastructure. Understanding and managing the flow of data across these interconnected ecosystems is crucial for preserving privacy.

***Regulatory Landscape:***

With the growing awareness of privacy issues, governments and regulatory bodies are enacting stringent laws and regulations to protect individuals' data rights. Navigating this evolving regulatory landscape adds an additional layer of complexity for organizations deploying IoT-based cloud solutions.

### 1.2 Objectives

This survey paper seeks to achieve several interrelated objectives, aiming to contribute to the existing body of knowledge on privacy in IoT-based cloud systems.

*Comprehensive Review:*

The primary objective is to conduct a comprehensive review of existing literature, research, and advancements in the realm of privacy-preserving techniques within the context of IoT-based cloud systems. By synthesizing diverse sources, we aim to provide a holistic understanding of the current state of the field.

*Classification of Privacy-Preserving Techniques:*

To facilitate a structured analysis, this survey paper will categorize and classify various privacy-preserving techniques. From encryption methods to access controls and AI-driven approaches, we aim to delineate the diverse strategies employed to safeguard data privacy in IoT-based cloud environments.

*In-Depth Analysis of Techniques:*

Each category of privacy-preserving techniques will undergo an in-depth analysis, unraveling the strengths, limitations, and potential applications. By scrutinizing the nuances of each approach, we seek to provide readers with insights into the practical considerations of implementing these techniques.

*Exploration of AI Integration:*

Recognizing the symbiotic relationship between artificial intelligence (AI) and privacy preservation, this survey paper will explore the integration of AI concepts within privacy-enhancing strategies. From machine learning for anonymization to AI-driven access control, we aim to highlight the transformative potential of AI in mitigating privacy risks.

*Case Studies and Real-World Applications:*

To ground theoretical discussions in practical contexts, this survey paper will incorporate case studies and examples of real-world implementations. By examining successful applications and challenges faced in different scenarios, readers can gain a nuanced understanding of the efficacy of privacy-preserving techniques.

*Comparative Analysis and Future Directions:*

A critical aspect of this survey paper involves conducting a comparative analysis of various privacy-preserving techniques. By evaluating their effectiveness, scalability, and adaptability, we aim to guide readers in selecting appropriate strategies based on specific use cases. Additionally, we will identify open challenges and propose potential directions for future research in the dynamic landscape of IoT-based cloud systems.

The groundwork for a comprehensive exploration of privacy in IoT-based cloud systems. As we navigate the intricate interplay between IoT, cloud computing, and data privacy, the subsequent sections will delve into the rich tapestry of privacy-preserving techniques, with a particular emphasis on the integration of artificial intelligence. By the journey's end, we anticipate not only shedding light on the current state of the field but also illuminating pathways for future research and innovation in this critical domain.

## 2. Literature Review

The literature review in this survey paper serves as a comprehensive examination of existing research on privacy issues within the realms of Internet of Things (IoT) and cloud computing. The objective is to distill key insights from the literature, identify challenges, discern emerging trends in privacy-preserving techniques, and categorize diverse approaches employed to secure data in IoT-based cloud systems.

### 2.1 Privacy Issues in IoT and Cloud Computing

The literature consistently highlights privacy concerns arising from the widespread deployment of IoT devices. Issues such as unauthorized data collection, inadequate user consent mechanisms, and the potential for data leakage have been recurrent themes. Research underscores the need for robust privacy measures to address the unique challenges posed by the vast and heterogeneous IoT ecosystem. The interaction between IoT and cloud computing amplifies privacy considerations [6-8]. The literature emphasizes that the centralization of data in cloud environments can lead to increased vulnerability. Issues like data access control, secure data transmission, and protection against insider threats are focal points of research, reflecting the growing recognition of cloud-related privacy challenges.

### 2.2 Key Challenges in Privacy-Preserving Techniques

*Scalability and Efficiency:*

One recurring challenge identified in the literature is the scalability of privacy-preserving techniques in the context of IoT-based cloud systems. As the number of connected devices continues to surge, ensuring that privacy measures can scale efficiently to handle the volume and diversity of data becomes paramount [9]. Efficient cryptographic protocols, distributed systems, and optimized algorithms are explored as potential solutions.

*Heterogeneity of IoT Devices:*

The diverse nature of IoT devices, ranging from resource-constrained sensors to powerful edge devices, introduces challenges in implementing uniform privacy-preserving measures [10]. The literature emphasizes the need for adaptive techniques that can cater to the varying computational capabilities, energy constraints, and communication protocols of different device classes.

*Interoperability and Standardization:*

Interoperability issues and the lack of standardized protocols across the IoT ecosystem are identified as significant challenges. The literature suggests that the absence of standardized security frameworks hampers the seamless integration of privacy-preserving techniques. Research in this area explores the development of interoperable solutions that can be universally applied across diverse IoT devices and platforms [11].

*Dynamic Nature of IoT Environments:*

The dynamic nature of IoT environments, where devices join or leave networks frequently, poses challenges in maintaining consistent privacy-preserving measures. The literature emphasizes the need for adaptive techniques that can seamlessly accommodate changes in device connectivity, ownership, and participation in the network.

**2.3 Trends in Privacy-Preserving Techniques**

*AI Integration for Enhanced Privacy:*

An emerging trend in the literature is the integration of artificial intelligence (AI) techniques to bolster privacy preservation. Machine learning, in particular, is explored for effective data anonymization in IoT data streams [12]. The use of AI-driven methods for generating synthetic data that preserves statistical properties while ensuring privacy is gaining traction. Additionally, AI is being employed in dynamic access control systems and anomaly detection for identifying unauthorized access attempts, showcasing its potential in enhancing privacy measures.

*Homomorphic Encryption for Secure Computation:*

Homomorphic encryption has emerged as a prominent trend in the literature, especially concerning secure computation in IoT-based cloud systems. The ability to perform computations on encrypted data without decryption is recognized as a powerful tool for preserving privacy during data transmission and storage. The literature explores the application of homomorphic encryption in tandem with AI, allowing machine learning algorithms to operate on encrypted data directly.

**2.4 Categorization of Privacy Preservation Approaches**

*Encryption Techniques:*

A substantial body of literature categorizes privacy-preserving techniques based on encryption. Homomorphic encryption, differential privacy, and attribute-based encryption are explored as methods to protect sensitive data during transmission, storage, and processing. The literature evaluates the strengths and limitations of each encryption technique, considering factors such as computational overhead and applicability to different use cases.

*Anonymization and Pseudonymization:*

Anonymization and pseudonymization techniques are extensively discussed in the literature as effective means to de-identify and protect individual identities in IoT data streams. The literature categorizes these techniques into static and dynamic approaches, with dynamic anonymization models adapting to changing data patterns.

*Access Control Mechanisms:*

Research in access control mechanisms focuses on dynamic and context-aware approaches. The literature categorizes access control techniques that leverage AI to adapt permissions based on user behavior, device characteristics, and contextual information. Behavioral analysis and anomaly detection are identified as integral components of AI-driven access control systems.

*Privacy-Preserving AI Integration:*

A notable category in the literature explores the integration of AI techniques for privacy preservation. Machine learning is employed for dynamic anonymization, generating synthetic data, and enhancing access control systems. The literature categorizes AI applications based on their specific roles in augmenting privacy-preserving measures.

## 3. IoT Architecture and Cloud Computing in Privacy Context

### 3.1 Typical Architecture of IoT-based Cloud Systems

Internet of Things (IoT)-based cloud systems encompass a complex and interconnected architecture designed to facilitate the seamless flow of data from IoT devices to the cloud for storage, processing, and analysis [13]. Understanding the typical architecture is paramount in identifying potential privacy concerns and implementing effective privacy-preserving measures.

*Sensors and Devices:*

At the forefront of the IoT ecosystem are sensors and devices embedded with connectivity features. These devices span a wide array, ranging from simple temperature sensors to sophisticated smart cameras and wearables. Each device plays a specific role in data collection, capturing information from the physical world.

*Edge Devices and Gateways:*

In many IoT scenarios, edge devices or gateways serve as intermediaries between sensors and the cloud [14]. These devices often pre-process data locally, reducing latency and bandwidth requirements. Edge computing plays a crucial role in offloading some computational tasks from the cloud, contributing to real-time responsiveness.

*Communication Protocols:*

Communication protocols act as the conduits through which data travels within the IoT architecture. Common protocols include MQTT (Message Queuing Telemetry Transport) and CoAP (Constrained Application Protocol). These lightweight and efficient protocols enable devices to transmit data to the cloud or edge seamlessly.

*Cloud Infrastructure:*

The heart of the IoT-based cloud system resides in the cloud infrastructure, typically composed of servers, databases, and scalable computing resources. Cloud platforms, such as Amazon Web Services (AWS), Microsoft Azure, and Google Cloud Platform (GCP), provide the necessary infrastructure for storing and processing vast amounts of data generated by IoT devices.

*Application Layer:*

On top of the cloud infrastructure, various applications and services are deployed to analyze and derive insights from the collected data. These applications range from simple dashboards for end-users to complex analytics engines and machine learning models for advanced data processing.

*Security Measures:*

Integrated into the architecture are security measures such as authentication, encryption, and access controls. These safeguards aim to protect data in transit and at rest, ensuring that only authorized entities can access and manipulate sensitive information.

**3.2 Data Flow through the IoT Ecosystem to the Cloud**

Understanding how data flows through the IoT ecosystem to the cloud is essential for identifying potential points of vulnerability and privacy risks.

*Data Generation at the Edge:*

The data generation process initiates at the edge, where sensors and devices capture information from the physical environment. This raw data can include anything from environmental metrics like temperature and humidity to personal health data collected by wearables.

*Local Processing and Edge Computing:*

In some cases, edge devices or gateways process data locally before transmitting it to the cloud. This local processing reduces the amount of raw data sent to the cloud, enhancing efficiency and reducing latency. Edge computing can involve data aggregation, filtering, or even running basic analytics on the device.

*Communication to the Cloud:*

Data is transmitted from devices to the cloud infrastructure through communication protocols. The choice of protocol depends on factors such as the nature of the data, latency requirements, and energy efficiency. During this transmission, data may pass through public or private networks, raising concerns about data interception and unauthorized access.

*Data Storage in the Cloud:*

Upon reaching the cloud infrastructure, data is stored in databases or data lakes. Cloud platforms offer scalable storage solutions, allowing organizations to accommodate the ever-growing volume of IoT-generated data. However, this centralization of data storage introduces potential privacy concerns related to data residency and compliance with regulatory frameworks.

*Data Processing and Analytics:*

The cloud's computational resources are leveraged for processing and analyzing the stored data [15]. This stage involves running algorithms, conducting analytics, and deriving meaningful insights from the collected information. While this processing enhances the value of IoT data, it also necessitates careful consideration of privacy implications, especially when dealing with sensitive personal or proprietary information.

*Insights and Actionable Outputs:*

The final stage involves presenting insights and actionable outputs derived from the processed data. This can manifest as visualizations, alerts, or automated actions triggered by specific data patterns. The delivery of these outputs raises privacy concerns, particularly regarding the dissemination of sensitive information and potential unintended consequences.

**3.3 Emphasizing Points Where Privacy Concerns May Arise**

Privacy concerns in IoT-based cloud systems are multifaceted, emerging at various stages of the data flow process. Identifying these points of vulnerability is crucial for implementing robust privacy-preserving measures.

***Data Collection and Sensor Vulnerabilities:***

The very initiation of data flow—data collection at the sensor level—poses privacy challenges. Sensors capturing personal or sensitive information may be vulnerable to physical tampering or compromise [16]. Additionally, the proliferation of IoT devices raises concerns about user consent and awareness regarding the types of data being collected.

***Edge Computing and Local Processing Risks:***

While edge computing offers advantages in terms of reduced latency and bandwidth usage, it introduces potential risks. Local processing may involve unsecured devices, making them susceptible to attacks. Ensuring the security of edge devices is paramount to prevent unauthorized access to processed or even raw data.

***Insecure Communication Protocols:***

Communication protocols play a pivotal role in data transmission. Insecure or poorly implemented protocols can expose data to interception and unauthorized access. Implementing secure communication channels, encryption, and authentication mechanisms are imperative to mitigate these risks.

***Cloud Storage and Data Residency:***

Storing data in the cloud introduces concerns related to data residency and compliance with regional or industry-specific regulations. Organizations must be cognizant of where their data is stored and ensure adherence to privacy laws governing the regions in which they operate.

***Data Processing and Analytics Challenges:***

The processing and analysis of data in the cloud demand careful consideration of privacy implications. Machine learning models trained on sensitive data may inadvertently reveal patterns that compromise individual privacy [17]. Anonymization and differential privacy techniques become essential to balance the benefits of data-driven insights with privacy preservation.

***Presentation of Insights and Automated Actions:***

The final stage, where insights are presented or automated actions are triggered, carries the risk of unintended consequences. If not carefully implemented, automated actions based on data patterns may lead to privacy violations or discriminatory outcomes. Transparency and explainability in automated decision-making processes become critical safeguards.

*Regulatory Compliance:*

Amidst the data flow, ensuring compliance with privacy regulations and standards is a pervasive concern [18]. The evolving landscape of data protection laws, such as the General Data Protection Regulation (GDPR) and the California Consumer Privacy Act (CCPA), necessitates constant vigilance to avoid legal ramifications and safeguard user rights.

The architecture and data flow of IoT-based cloud systems illuminates the various stages where privacy concerns may arise. Addressing these concerns requires a holistic approach that integrates security measures, privacy-preserving techniques, and adherence to regulatory frameworks. The subsequent sections of this survey paper will delve into the diverse privacy-preserving strategies employed across these stages, with a specific focus on innovative approaches and the integration of artificial intelligence for enhanced privacy protection.

## 4. Privacy Threats in IoT-based Cloud Systems

### 4.1 Common Privacy Threats in IoT Data Collection

The advent of Internet of Things (IoT) devices has ushered in a paradigm shift in how we interact with and perceive our surroundings. However, this proliferation of connected devices brings along a host of privacy threats, especially concerning the sensitive data they collect.

*Unauthorized Data Collection:*

One of the primary threats stems from unauthorized or undisclosed data collection. IoT devices, ranging from smart home appliances to wearable health trackers, often collect more data than users are aware of. This can include personally identifiable information (PII), behavioral patterns, and even audio or video recordings, raising concerns about user consent and privacy infringement.

*Inadequate Security Measures:*

IoT devices are frequently characterized by resource constraints, making them susceptible to security vulnerabilities [19]. Insecure communication channels, lack of encryption, and insufficient authentication mechanisms can expose data to unauthorized access. Malicious actors may exploit these vulnerabilities to intercept or manipulate the data in transit.

*Lack of Standardized Security Protocols:*

The absence of standardized security protocols across all IoT devices exacerbates the challenge of ensuring robust security. Each device may operate on different communication standards and security measures, making it challenging to enforce a unified security framework. This lack of standardization increases the attack surface and weakens overall security posture.

*Insecure Device Firmware and Software:*

IoT devices often run on firmware and software that may have vulnerabilities [20]. Manufacturers may not prioritize regular updates or security patches, leaving devices exposed to exploits. Attackers can leverage these vulnerabilities to gain unauthorized access,

*Data Interception and Eavesdropping:*

The communication between IoT devices and the cloud is susceptible to interception and eavesdropping. If communication channels lack encryption, adversaries can capture and analyze the data in transit. This interception can reveal sensitive information, posing a significant threat to user privacy.

**4.2 Risks Associated with Storing and Processing IoT Data in the Cloud**

Once data is collected by IoT devices, it is transmitted to the cloud for storage and processing. This phase introduces a new set of privacy risks and challenges that organizations must address to safeguard user information.

*Data Residency and Jurisdictional Issues:*

Storing IoT data in the cloud may involve transferring data across geographic regions [21]. Different countries have varying data protection laws, and the lack of careful consideration regarding data residency can lead to legal ramifications. Organizations must navigate these jurisdictional challenges to ensure compliance with relevant regulations and protect user privacy.

*Insufficient Access Controls:*

In a cloud environment, data accessibility is a critical aspect of privacy protection. Inadequate access controls may allow unauthorized personnel or entities to access sensitive IoT data. Implementing robust access controls, encryption, and authentication mechanisms is essential to prevent unauthorized access and data breaches.

*Cloud Service Provider Vulnerabilities:*

Cloud service providers host and manage vast amounts of IoT data. However, these providers are not immune to security vulnerabilities. Breaches or vulnerabilities in the cloud infrastructure can lead to unauthorized access to stored data. Organizations must choose reputable and secure cloud providers while also implementing additional security measures to mitigate these risks.

*Insider Threats and Misuse of Privileges:*

The cloud environment introduces the possibility of insider threats, where individuals with authorized access misuse their privileges. This can range from accidental data exposure to intentional data manipulation. Organizations need to monitor and manage user privileges effectively, implementing measures to detect and prevent insider threats.

*Data Silos and Integration Challenges:*

IoT data stored in the cloud may exist in silos, making it challenging to integrate and analyze data comprehensively. Siloed data increases the risk of incomplete or fragmented insights, potentially leading to privacy issues if critical patterns or relationships are overlooked. Establishing effective data integration strategies is crucial for a holistic understanding of the information collected.

*Data Retention and Deletion Challenges:*

Organizations may retain IoT data for extended periods, posing risks to user privacy. The longer data is stored, the higher the likelihood of it being subject to unauthorized access or use. Establishing clear data retention policies and mechanisms for secure data deletion is imperative to mitigate these risks and align with privacy principles.

*Inadequate Data Anonymization:*

While processing IoT data in the cloud, organizations must prioritize data anonymization to protect user privacy. Incomplete or ineffective anonymization techniques can lead to the unintentional disclosure of sensitive information. Striking a balance between meaningful analysis and preserving individual privacy requires advanced anonymization strategies.

### 4.3 Mitigation Strategies for Privacy Threats

Effectively addressing privacy threats in IoT-based cloud systems requires a proactive and multi-faceted approach. Implementing robust security measures and adopting privacy-preserving technologies can significantly mitigate these threats.

*Security by Design:*

Incorporate security measures into the design and development of IoT devices. Prioritize secure coding practices, conduct regular security assessments, and ensure that firmware and software updates are part of the device lifecycle.

*Encryption and Secure Communication:*

Implement end-to-end encryption for data transmitted between IoT devices and the cloud. Secure communication protocols, such as TLS (Transport Layer Security), can prevent unauthorized access and eavesdropping.

*Standardization of Security Protocols:*

Advocate for standardized security protocols across the IoT ecosystem. Establishing uniform security standards can enhance interoperability, simplify security management, and create a more secure environment for IoT devices.

*Regular Security Audits and Updates:*

Conduct regular security audits of IoT devices and cloud infrastructure. Promptly address identified vulnerabilities through software updates and patches. Regularly assess and enhance security measures to adapt to evolving threats.

*Privacy Impact Assessments:*

Perform privacy impact assessments (PIAs) before deploying IoT-based cloud systems. Evaluate the potential privacy risks associated with data collection, storage, and processing. Implement measures to address identified risks and ensure compliance with privacy regulations.

*Granular Access Controls:*

Implement granular access controls in the cloud environment. Define and enforce access policies based on the principle of least privilege, ensuring that only authorized individuals or systems can access specific data.

*Choose Secure Cloud Service Providers:*

Select reputable and secure cloud service providers with a strong track record of data security. Evaluate the provider's security practices, certifications, and compliance with industry standards.

*Data Anonymization Techniques:*

Adopt advanced data anonymization techniques during data processing in the cloud. Ensure that personally identifiable information is effectively anonymized to prevent the identification of individuals while still allowing for meaningful analysis.

*Transparent Data Usage Policies:*

Establish clear and transparent data usage policies for IoT-based cloud systems. Inform users about the types of data collected, the purposes for which it will be used, and how long it will be retained. Obtain explicit consent for data processing activities.

*Periodic Privacy Training:*

Provide periodic privacy training for individuals involved in the development, deployment, and maintenance of IoT-based cloud systems. Educate users about privacy best practices and empower them to make informed decisions regarding their data.

## 5. Privacy-Preserving Techniques

Privacy-preserving techniques play a pivotal role in mitigating the risks associated with the collection, storage, and processing of sensitive data in IoT-based cloud systems [22]. These techniques encompass a spectrum of strategies, ranging from encryption and anonymization to access control and differential privacy.

### 5.1 Classification of Privacy-Preserving Techniques

#### 5.1.1 Encryption

Encryption is a fundamental privacy-preserving technique that involves converting plaintext data into ciphertext using cryptographic algorithms. It ensures that only authorized entities with the appropriate decryption keys can access the original data.

**Categories:**

Symmetric Encryption: Uses a single key for both encryption and decryption.

Asymmetric Encryption: Involves a pair of public and private keys for encryption and decryption, respectively.

#### 5.1.2 Anonymization

Anonymization aims to transform or remove personally identifiable information (PII) from datasets, making it challenging to identify individuals [23]. Common techniques include generalization, suppression, and perturbation.

**Categories:**

K-Anonymity: Ensures that each record in a dataset is indistinguishable from at least k-1 other records.

Differential Privacy: Quantifies the impact of an individual's data on the overall result of a query, providing strong privacy guarantees.

### 5.1.3 Access Control

Access control regulates who can access specific data and what actions they can perform. It involves authentication mechanisms, authorization policies, and user permissions.

**Categories**:

Role-Based Access Control (RBAC): Assigns permissions based on predefined roles.

Attribute-Based Access Control (ABAC): Grants access based on attributes associated with users, resources, and the environment.

### 5.1.4 Homomorphic Encryption

Homomorphic encryption allows computations to be performed on encrypted data without decrypting it. This technique enables secure processing of sensitive information in the encrypted state.

**Categories**:

Partially Homomorphic Encryption: Supports only one operation (e.g., addition or multiplication) on encrypted data.

Fully Homomorphic Encryption: Supports both addition and multiplication operations on encrypted data.

### 5.1.5 Federated Learning

Federated learning enables model training across decentralized devices without exchanging raw data. Instead, only model updates are shared, preserving individual data privacy.

**Categories**:

Horizontal Federated Learning: Shares data samples across devices.

Vertical Federated Learning: Shares features or attributes across devices.

### 5.1.6 Differential Privacy

Differential privacy adds noise to data before analysis to prevent the identification of individual records [24]. It aims to ensure that the presence or absence of a specific record does not significantly impact the outcome of a computation.

**Categories:**

Central Differential Privacy: The noise is added centrally before data analysis.

Local Differential Privacy: Each participant adds noise to their data before sharing it with the central server.

### 5.2 In-Depth Analysis of Each Technique

#### 5.2.1 Encryption

**Strengths:**

- Provides strong data confidentiality.
- Effective in preventing unauthorized access during data transmission and storage.
- Supports secure computation on encrypted data.

**Limitations:**

- Introduces computational overhead during encryption and decryption processes.
- Key management can be challenging, especially in large-scale systems.
- Does not address issues related to metadata leakage.

#### 5.2.2 Anonymization

**Strengths:**

- Protects individual privacy by obscuring personally identifiable information.
- Facilitates the sharing of data for analysis without revealing sensitive details.
- Can be applied to various data types and formats.

**Limitations:**

- Risk of re-identification, especially with the availability of auxiliary information.
- Aggressive anonymization may lead to a loss of data utility.
- Balancing anonymity and data quality can be challenging.

#### 5.2.3 Access Control

**Strengths:**

- Provides a granular level of control over data access.
- Helps in preventing unauthorized users from accessing sensitive information.
- Facilitates compliance with regulatory requirements regarding data access.

**Limitations:**

- Complex to manage in large and dynamic systems.
- May lead to administrative overhead in defining and updating access policies.
- Access control mechanisms can be vulnerable to exploitation if not properly implemented.

#### 5.2.4 Homomorphic Encryption

**Strengths:**

- Enables secure computation on encrypted data, preserving privacy.
- Allows for secure outsourcing of data processing tasks to third-party providers.
- Supports privacy-preserving machine learning applications.

**Limitations:**

- Introduces computational complexity and performance overhead.
- Current implementations may not be as efficient as non-encrypted computations.
- Limited support for certain types of operations compared to traditional computing.

### 5.2.5 Federated Learning

**Strengths:**

- Preserves individual data privacy by keeping data on local devices.
- Facilitates collaborative model training without sharing raw data.
- Suitable for decentralized environments and edge computing scenarios.

**Limitations:**

- Synchronization challenges when aggregating model updates from diverse devices.
- Potential security risks if model updates are intercepted or manipulated.
- May require more communication rounds for convergence compared to centralized approaches.

### 5.2.6 Differential Privacy

**Strengths:**

- Offers strong privacy guarantees by preventing the identification of individual contributions.
- Can be applied to a wide range of data analysis tasks.
- Provides a quantifiable measure of privacy protection.

**Limitations:**

- Introduces noise, affecting the accuracy of query results.
- Balancing privacy and data utility is a delicate trade-off.
- Challenges in determining appropriate noise levels for different queries.

## 5.3 Recent Advancements in Privacy-Preserving Technologies

### 5.3.1 Homomorphic Encryption Advances

Recent advancements in homomorphic encryption focus on improving efficiency and expanding the range of supported operations [25-27]. Innovations such as lattice-based cryptography and optimized algorithms aim to reduce computational overhead, making homomorphic encryption more practical for real-world applications.

### 5.3.2 Federated Learning Innovations

Research in federated learning explores techniques to enhance communication efficiency and model convergence. Federated transfer learning, secure aggregation protocols, and adaptive client selection algorithms are among the advancements aimed at addressing challenges associated with federated learning in diverse and dynamic environments.

### 5.3.3 Differential Privacy Developments

In the realm of differential privacy, ongoing research focuses on refining noise injection mechanisms to achieve a better balance between privacy and data utility [28]. Improved privacy amplification techniques, adaptive privacy budgets, and context-aware differential privacy are emerging as key areas of innovation.

### 5.3.4 Integration of Multiple Techniques

Recent trends indicate a growing emphasis on integrating multiple privacy-preserving techniques to address diverse threats comprehensively. For example, combining homomorphic encryption with federated learning allows for secure and privacy-preserving model training on encrypted data.

### 5.3.5 User-Centric Privacy Solutions

Advancements in user-centric privacy solutions emphasize empowering individuals to have more control over their data. Self-sovereign identity systems, decentralized identity platforms, and user-centric data sharing protocols.

## 5.4 Privacy-Preserving Techniques with AI Integration

The integration of artificial intelligence (AI) with privacy-preserving techniques is a dynamic and evolving field, particularly in the context of IoT-based cloud systems [29]. This section delves into the synergy between AI and privacy preservation, exploring how machine learning can enhance data anonymization, the role of homomorphic encryption in conjunction with AI, and the application of AI in adaptive access control systems.

### 5.4.1 Machine Learning for Anonymization

*Challenges in Traditional Anonymization:*

Traditional anonymization techniques, such as generalization and suppression, face challenges in handling the dynamic and heterogeneous nature of IoT data streams. The sheer volume and variability of data generated by diverse IoT devices necessitate more adaptive and intelligent approaches.

*Machine Learning Approaches:*

Machine learning techniques offer a promising avenue for effective data anonymization in IoT data streams. These approaches leverage the learning capabilities of algorithms to understand and adapt to the unique characteristics of data over time.

*Dynamic Anonymization Models:*

Machine learning models can be trained to dynamically adjust anonymization parameters based on the evolving data patterns. For instance, clustering algorithms can group similar data points together, allowing for anonymization strategies that preserve the statistical properties of each cluster while ensuring privacy.

*AI-Driven Contextual Anonymization:*

Contextual anonymization, where the level of anonymization is determined by the context of the data, can be achieved through AI-driven models [30]. These models learn from contextual cues in the data stream to apply appropriate anonymization techniques, balancing the need for privacy with data utility.

### 5.4.2 AI-Generated Synthetic Data for Privacy Preservation

*Synthetic Data Generation:*

AI plays a pivotal role in generating synthetic data that mimics the statistical properties of real data while preserving individual privacy. Generative models, such as Generative Adversarial Networks (GANs) and Variational Autoencoders (VAEs), can create synthetic datasets that are indistinguishable from the original, yet devoid of personally identifiable information.

*Preserving Data Distribution:*

Machine learning models can be trained to understand and replicate the underlying distribution of sensitive features in the data [31]. This ensures that synthetic data maintains the same statistical properties as the original data, crucial for preserving the utility of the data for analysis.

*Differential Privacy and AI-Generated Data:*

Integrating differential privacy principles with AI-generated data adds an extra layer of privacy protection. AI models can be trained with differential privacy techniques to inject controlled noise into the synthetic data, further safeguarding against potential privacy breaches.

### 5.5 Homomorphic Encryption and AI

### 5.5.1 Securing Data in Transit and at Rest with Homomorphic Encryption

Homomorphic encryption allows computations to be performed on encrypted data without the need for decryption. This property makes it particularly valuable for securing data in transit and at rest in IoT-based cloud systems.

*Role in Data Transmission:*

In IoT scenarios, where data is transmitted between devices and cloud servers, homomorphic encryption ensures end-to-end security [32]. AI-driven applications can leverage this encrypted data without exposing sensitive information during transmission, providing a robust layer of privacy.

*Securing Data at Rest:*

When data is stored in the cloud or on edge devices, homomorphic encryption prevents unauthorized access. Even when AI algorithms need to access and process this stored data, they can do so without decrypting it, maintaining the confidentiality of the information.

### 5.5.2 AI Operating on Encrypted Data

*Preserving Privacy during Computation:*

One of the significant advantages of homomorphic encryption is that it allows AI algorithms to operate on encrypted data directly. This means that machine learning models can perform computations, train on encrypted datasets, and generate insights without ever exposing the raw, unencrypted data.

*Use Cases:*

In practical terms, this capability is beneficial for scenarios where data privacy is paramount. For example, in healthcare IoT applications, where sensitive patient data is involved, AI algorithms can perform analyses on encrypted medical records without compromising individual privacy.

*Challenges and Considerations:*

While the prospect of AI operating on encrypted data is promising for privacy preservation, it comes with challenges [33]. Homomorphic encryption introduces computational overhead, and the choice of encryption schemes and parameters must be carefully considered to balance security and performance.

## 5.6 AI-Driven Access Control

### 5.6.1 Dynamic Access Control Systems with AI

*Adaptability of Access Control:*

Traditional access control systems often struggle to adapt to the dynamic nature of IoT environments. AI-driven access control systems bring a level of adaptability that is crucial in scenarios where devices join or leave networks frequently.

*Behavioral Analysis for Access Control:*

AI algorithms can analyze user and device behavior patterns to dynamically adjust access permissions. For instance, a device that exhibits anomalous behavior may have its access restricted, preventing potential security breaches and ensuring privacy.

*Context-Aware Access Control:*

Context-aware access control involves considering contextual information, such as the time of day, location, and the specific device's historical behavior, to make access decisions [34]. AI models can learn and predict access requirements based on contextual cues, enhancing the precision of access control.

### 5.6.2 AI-Based Anomaly Detection for Unauthorized Access

*Importance of Anomaly Detection:*

Identifying unauthorized access attempts is crucial for maintaining the integrity of privacy-preserving systems. Traditional rule-based methods may fall short in identifying sophisticated attacks or subtle anomalies in user behavior.

*Machine Learning for Anomaly Detection:*

AI-driven anomaly detection employs machine learning models trained on historical data to recognize patterns associated with normal behavior [35]. Any deviation from these patterns triggers alerts, signaling potential unauthorized access attempts.

*Continuous Learning Models:*

AI-driven anomaly detection systems can adapt and learn from new data, ensuring that the model evolves to recognize emerging threats. This continuous learning approach enhances the robustness of access control mechanisms over time.

### 5.7 Considerations and Challenges

#### 5.7.1 Ethical Considerations in AI Integration

*Responsible AI Use:*

Integrating AI into privacy-preserving systems necessitates a commitment to ethical considerations. Ensuring that AI models are used responsibly and do not perpetuate biases or compromise fairness is paramount.

*Transparency and Explainability:*

AI algorithms must be transparent and explainable, particularly when handling sensitive data. Users and stakeholders should be able to understand the decision-making processes of AI-driven systems, fostering trust and accountability.

#### 5.7.2 Computational Overhead and Efficiency

*Balancing Security and Performance:*

Homomorphic encryption introduces computational overhead, and AI algorithms operating on encrypted data may experience reduced efficiency [36]. Striking a balance between security and performance is critical, and optimizing algorithms and encryption schemes becomes imperative.

#### 5.7.3 Data Quality and Utility

*Preserving Data Utility:*

While privacy is a top priority, it's essential to preserve the utility of data for meaningful analysis. AI-driven approaches, whether in anonymization or synthetic data generation, should be designed to maintain the quality and relevance of the data for intended applications.

#### 5.7.4 Privacy-Preserving Regulations

*Alignment with Regulations:*

AI integration in privacy preservation should align with existing and emerging privacy regulations. Ensuring compliance with standards and regulations is crucial for the ethical and legal use of AI-driven techniques in IoT-based cloud systems.

## 6. Case Studies with AI Applications in Privacy Preservation

The application of artificial intelligence (AI) in privacy preservation extends beyond traditional methods. This section explores case studies where AI algorithms play a pivotal role in smart AI-powered privacy policies and predictive privacy maintenance. By analyzing privacy policies and predicting potential privacy breaches, these AI-driven approaches exemplify innovative solutions in safeguarding data and ensuring compliance with regulations and standards.

### 6.1 Smart AI-Powered Privacy Policies

Privacy policies and terms of service are integral components of any digital platform, outlining how user data is collected, processed, and shared. Analyzing and ensuring compliance with these policies can be a daunting task, especially as the complexity and legal nuances evolve. AI provides a solution by automating the analysis of privacy policies to ensure alignment with regulations and standards [37-39]. A leading e-commerce platform implemented an AI-driven compliance checker to analyze and assess the privacy policies and terms of service of third-party vendors and partners. The platform aimed to ensure that all entities in its ecosystem adhered to global privacy regulations, including GDPR and CCPA. The AI model was trained on a diverse dataset of privacy policies and legal documents, learning to identify clauses, language patterns, and key terms associated with privacy compliance [40]. Natural Language Processing (NLP) techniques were employed to extract and categorize relevant information from textual documents.

#### 6.1.1 *Key Functionalities:*

Regulatory Alignment: The AI model flagged sections of privacy policies that required attention, ensuring alignment with specific regulatory requirements.

Policy Changes Tracking: Continuous monitoring allowed the system to track changes in privacy policies over time, alerting the platform to updates that might impact compliance.

Vendor Risk Assessment: The AI-driven system assessed the risk associated with each vendor or partner based on their privacy policies, aiding in risk management and decision-making.

#### 6.1.2 Benefits:

Efficiency: The automation of compliance checks significantly reduced the time and resources required for manual policy analysis.

Proactive Compliance: The system enabled proactive identification of compliance gaps, allowing the platform to address issues before they became critical.

Scalability: As the ecosystem expanded, the AI-driven compliance checker easily scaled to accommodate a growing number of partners and their evolving privacy policies.

### 6.1.3 Future Directions

**Enhanced Regulatory Coverage:**

Future developments in this AI application could involve expanding the model's coverage to new and evolving privacy regulations globally. This ensures that the platform stays ahead of regulatory changes and remains adaptable to the dynamic legal landscape.

User-Focused Privacy Evaluation:

AI algorithms could be fine-tuned to evaluate privacy policies from a user-centric perspective. This involves assessing how comprehensible and transparent the policies are to users, enhancing overall user empowerment and trust.

**Cross-Platform Integration:**

To create a more comprehensive privacy ecosystem, the AI-driven compliance checker could evolve to integrate with other platforms, fostering industry-wide collaboration and standardization in privacy practices.

### 6.2 Predictive Privacy Maintenance

Predictive privacy maintenance leverages AI models to anticipate and prevent potential privacy breaches before they occur. By analyzing historical data, user behavior, and system vulnerabilities, these models proactively identify and mitigate risks, ensuring a robust privacy posture [41]. A healthcare organization adopted an AI-driven predictive privacy maintenance system to safeguard patient data and maintain compliance with health information privacy laws. The system aimed to prevent unauthorized access, data leaks, and other potential privacy breaches. The predictive model integrated machine learning algorithms to analyze patterns in user access, data usage, and system vulnerabilities. It employed anomaly detection techniques to identify deviations from normal behavior and predict potential privacy risks.

#### 6.2.1 *Key Functionalities:*

User Behavior Analysis: The AI model continuously analyzed user access patterns, distinguishing between normal and anomalous behavior.

Data Flow Monitoring: Real-time monitoring of data flow within the system allowed the model to track and predict potential data leaks or unauthorized transfers.

Vulnerability Assessment: By incorporating information on system vulnerabilities and patch levels, the model predicted possible entry points for privacy breaches.

#### 6.2.2 *Benefits:*

Proactive Risk Mitigation: Predictive analysis enabled the organization to address potential privacy risks before they manifested, reducing the likelihood of data breaches.

Adaptive Security Measures: The system autonomously adapted security measures based on evolving user behavior and emerging threats.

Compliance Assurance: By continuously monitoring and predicting privacy risks, the organization maintained a high level of compliance with healthcare privacy regulations.

### 6.2.3 *Future Directions*

*Behavioral Biometrics Integration:*

Future iterations of predictive privacy maintenance systems could incorporate behavioral biometrics, such as keystroke dynamics and mouse movement patterns. This would add an additional layer of user authentication and anomaly detection.

*Explainable AI for Privacy Predictions:*

To enhance transparency and trust, the AI models could be designed to provide explanations for their predictions. This ensures that stakeholders, including privacy officers and regulatory bodies, can understand the rationale behind predicted privacy risks.

*Cross-Sector Application:*

The principles of predictive privacy maintenance are applicable across various sectors. Future developments might involve adapting and refining these models for industries beyond healthcare, such as finance and telecommunications.

## 6.3 Considerations and Challenges

### 6.3.1 Ethical Implications

*User Consent and Transparency:*

Both AI-powered privacy policies and predictive privacy maintenance raise ethical considerations related to user consent and transparency. Organizations must ensure that users are informed about the AI-driven processes and provide clear mechanisms for opting in or out.

*Avoiding Bias in Predictive Models:*

Predictive privacy maintenance relies heavily on historical data, which may introduce biases. It is crucial to implement measures to identify and mitigate biases to ensure fair and unbiased predictions.

### 6.3.2 Technological Challenges

*Integration Complexity:*

Implementing AI-driven privacy solutions may pose integration challenges with existing systems. Ensuring seamless integration without disrupting regular operations is a critical consideration.

*Data Security and Privacy:*

While these AI applications focus on preserving user privacy, it's essential to address potential risks related to the security of the AI models themselves. Ensuring the confidentiality and integrity of AI models and data inputs is paramount.

The case studies presented demonstrate the transformative impact of integrating AI into privacy preservation strategies. Whether in the form of smart AI-powered privacy policies or predictive privacy maintenance, these applications showcase the ability of AI to enhance compliance, proactively manage risks, and create a more resilient and privacy-aware digital ecosystem. As AI continues to evolve, organizations must navigate ethical considerations and technological challenges to unlock the full potential of these innovative privacy-preserving solutions.

## 7. Comparative Analysis of Privacy-Preserving Techniques

Privacy-preserving techniques are integral components of securing sensitive data in IoT-based cloud systems. Each technique has its unique strengths and weaknesses, making them suitable for different scenarios and applications [42]. A comparative analysis enables us to understand the nuanced aspects of these techniques, evaluate their effectiveness, and explore the trade-offs inherent in their implementation.

### 7.1 Comparison of Privacy-Preserving Techniques

#### 7.1.1 Encryption

Data Confidentiality: Encryption provides a robust solution for maintaining the confidentiality of data, ensuring that only authorized entities can access the original information.

Versatility: Encryption can be applied at various stages, including data in transit, data at rest, and even during computation, offering a versatile approach to privacy preservation.

*Weaknesses:*

Computational Overhead: The encryption and decryption processes can introduce computational overhead, especially in resource-constrained environments typical of IoT devices.

Metadata Leakage: While encryption protects the content of the data, metadata such as the size and frequency of encrypted messages may still be susceptible to analysis.

#### 7.1.2 Anonymization

User Privacy: Anonymization protects user privacy by transforming or removing personally identifiable information (PII), enabling data sharing for analysis without compromising individual identities.

Applicability: Anonymization techniques can be applied to various data types, making them versatile for scenarios with diverse data sources.

*Weaknesses:*

Risk of Re-Identification: Aggressive anonymization may not always guarantee protection against re-identification, especially when auxiliary information is available.

Data Utility Trade-Off: There is often a trade-off between data utility and anonymity, where extreme anonymization may lead to a loss of valuable information.

### 7.1.3 Access Control

Granular Control: Access control provides granular control over who can access specific data, ensuring that only authorized users or entities have the necessary permissions.

Regulatory Compliance: Access control mechanisms facilitate compliance with regulatory requirements regarding data access and user permissions.

*Weaknesses:*

Complexity: Managing access control in large and dynamic systems can be complex, requiring careful definition and continuous updates of access policies.

Potential for Exploitation: If not properly implemented, access control mechanisms can be vulnerable to exploitation, leading to unauthorized access and potential privacy breaches.

### 7.1.4 Homomorphic Encryption

Secure Computation: Homomorphic encryption enables secure computation on encrypted data, preserving privacy during processing.

Outsourcing Computations: It allows organizations to outsource data processing tasks to third-party providers while maintaining the confidentiality of the data.

*Weaknesses:*

Computational Complexity: Homomorphic encryption introduces computational complexity, impacting the performance of data processing tasks.

Limited Operations: Certain operations may be less efficient with homomorphic encryption compared to traditional computations, limiting its applicability in certain scenarios.

### 7.1.5 Federated Learning

Decentralized Data: Federated learning operates on decentralized data, preserving individual privacy by keeping sensitive information on local devices.

Collaborative Model Training: It facilitates collaborative model training without sharing raw data, making it suitable for scenarios where data cannot be centralized.

*Weaknesses:*

Communication Overhead: Federated learning introduces communication challenges, requiring synchronization of model updates from diverse devices.

Security Risks: If model updates are intercepted or manipulated during the aggregation process, it can pose security risks to the privacy of the data.

### 7.1.6 Differential Privacy

Quantifiable Privacy Guarantees: Differential privacy provides quantifiable privacy guarantees, offering a measurable threshold for the impact of individual data contributions.

Versatility: It can be applied to various data analysis tasks, providing a flexible approach to privacy preservation.

*Weaknesses:*

Noise Impact on Accuracy: Introducing noise to achieve differential privacy can impact the accuracy of query results, requiring careful tuning to balance privacy and data utility.

Trade-Off Challenges: Achieving an optimal balance between privacy and data utility often involves complex trade-offs, and determining appropriate noise levels for different queries can be challenging.

**7.2 Evaluation of Effectiveness in Different Scenarios**

**7.2.1 Data Sensitivity and Criticality**

*Encryption:*

Effectiveness: Encryption is highly effective for scenarios where data sensitivity and criticality are paramount. It ensures that even if unauthorized access occurs, the encrypted data remains indecipherable.

Considerations: The choice of encryption algorithms and key management practices becomes crucial in determining the overall security posture.

*Anonymization:*

Effectiveness: Anonymization is suitable for scenarios where individual privacy is a primary concern [43]. It is often employed in scenarios where aggregated insights are valuable, and individual identities need to be protected.

Considerations: The level of anonymization applied should align with the sensitivity of the data and the intended use of the anonymized dataset.

*Access Control:*

Effectiveness: Access control is effective in scenarios where there is a need for granular control over data access. It is particularly valuable in environments where specific individuals or entities require varying levels of access.

Considerations: Regularly updating access policies and ensuring strong authentication mechanisms are critical considerations.

*Homomorphic Encryption:*

Effectiveness: Homomorphic encryption is well-suited for scenarios where secure computation on encrypted data is essential. It finds applications in situations where privacy-preserving data processing is critical.

Considerations: The computational overhead introduced by homomorphic encryption should be considered in resource-constrained environments.

# 8. Open Challenges and Future Directions in Privacy Preservation for IoT-based Cloud Systems

## 8.1 Current Gaps and Challenges

### 8.1.1 Scalability Issues

The exponential growth of IoT devices contributes to scalability challenges in privacy preservation. As the number of connected devices increases, traditional privacy-preserving techniques may struggle to scale efficiently, leading to computational bottlenecks and performance degradation. Developing scalable and efficient privacy-preserving solutions that can handle the growing volume and diversity of IoT-generated data is a critical challenge. This includes exploring distributed and parallelized approaches to ensure scalability without compromising on privacy.

### 8.1.2 Heterogeneity in IoT Ecosystems

The heterogeneity of IoT devices, ranging from resource-constrained sensors to powerful edge devices, poses challenges in implementing uniform privacy-preserving measures. One-size-fits-all solutions may not be practical due to the diverse nature of devices and their capabilities. Research is needed to devise adaptive privacy-preserving techniques that can cater to the varying computational capabilities, energy constraints, and communication protocols of different IoT devices. Tailoring solutions to the specific characteristics of each device class is crucial for effective privacy preservation.

### 8.1.3 Interoperability and Standardization

The lack of standardized protocols and interoperability standards in the IoT ecosystem hinders the seamless integration of privacy-preserving techniques. Devices and platforms from different manufacturers may operate on disparate security frameworks, complicating the implementation of unified privacy measures. Fostering industry-wide collaboration to establish standardized security protocols and interoperability standards is essential. Future research should focus on developing frameworks that enable the seamless integration of privacy-preserving solutions across diverse IoT devices and platforms.

### 8.1.4 Dynamic Nature of IoT Environments

IoT environments are dynamic, with devices joining or leaving networks frequently. This dynamic nature introduces challenges in maintaining consistent privacy-preserving measures, especially when dealing with access control and key management. Addressing the dynamic nature of IoT environments requires the development of adaptive privacy-preserving techniques that can seamlessly accommodate changes in device connectivity, ownership, and participation in the network. Dynamic key management and access control mechanisms are areas that merit focused research.

### 8.1.5 User Awareness and Control

End-users often lack awareness of the privacy implications associated with IoT devices and the data they generate. Moreover, users may have limited control over how their data is collected, processed, and shared by these devices. Enhancing user awareness and empowering individuals with greater control over their data are critical challenges. Future research should explore user-centric privacy solutions, including intuitive interfaces, transparent data usage policies, and mechanisms for users to exert more control over their data.

### 8.1.6 Privacy-Preserving AI Integration

As AI plays an increasingly prominent role in processing and deriving insights from IoT data, ensuring privacy-preserving AI integration becomes a complex challenge. Machine learning models trained on sensitive data may inadvertently leak information or perpetuate biases. Developing privacy-preserving AI models, federated learning approaches, and differential privacy techniques that can be seamlessly integrated into IoT-based cloud systems is a pressing challenge. This involves mitigating privacy risks associated with AI model inference and ensuring fairness and transparency in AI-driven decisions.

## 8.2 Potential Avenues for Future Research

### 8.2.1 Hybrid Privacy-Preserving Models

*Future Direction:*

Research should explore the development of hybrid privacy-preserving models that combine multiple techniques synergistically. For example, integrating homomorphic encryption with federated learning or combining anonymization with differential privacy could offer enhanced privacy guarantees while addressing specific challenges associated with each technique.

*Rationale:*

Hybrid models can leverage the strengths of individual techniques to mitigate their respective weaknesses. This approach may lead to more robust and adaptable privacy-preserving solutions that cater to the multifaceted nature of IoT-based cloud systems.

### 8.2.2 Explainable and Auditable Privacy

*Future Direction:*

There is a growing need for privacy-preserving solutions that are not only effective but also explainable and auditable. Future research should focus on developing techniques that provide transparency into how privacy is preserved, enabling users and stakeholders to understand and audit the privacy measures in place.

*Rationale:*

Explainability and auditability are crucial for building trust in privacy-preserving systems. Users, regulators, and organizations should be able to comprehend the mechanisms employed for privacy protection, fostering transparency and accountability in IoT-based cloud environments.

### 8.2.3 Edge-Driven Privacy Preservation

*Future Direction:*

With the increasing prevalence of edge computing in IoT scenarios, future research should investigate edge-driven privacy-preserving techniques. This involves moving privacy-preserving computations closer to the data source, minimizing data transmission to centralized cloud servers.

*Rationale:*

Edge-driven privacy preservation can alleviate bandwidth constraints, reduce latency, and enhance overall system efficiency. Exploring how edge devices can autonomously enforce privacy measures without relying heavily on central cloud infrastructure is a promising avenue.

### 8.2.4 Federated Learning Enhancements

*Future Direction:*

Advancements in federated learning should focus on overcoming existing challenges related to communication overhead, security risks, and model convergence. Future research may explore novel algorithms, adaptive federated learning approaches, and privacy-preserving techniques specifically designed for decentralized IoT environments.

*Rationale:*

Enhanced federated learning methodologies can significantly contribute to privacy preservation in scenarios where data decentralization is a fundamental requirement. Overcoming communication challenges and ensuring robust security mechanisms will be crucial for the success of federated learning in IoT.

### 8.2.5 Privacy-Preserving Regulations and Standards

*Future Direction:*

Research efforts should also be directed towards the development of privacy-preserving regulations and standards specific to IoT-based cloud systems. This involves collaborating with regulatory bodies, industry stakeholders, and privacy experts to establish guidelines that ensure the consistent application of privacy measures across IoT environments.

*Rationale:*

Standardized regulations and guidelines can provide a framework for organizations to adhere to and can serve as a benchmark for assessing the adequacy of privacy-preserving measures. This is particularly important in the absence of comprehensive and domain-specific privacy regulations for IoT.

### 8.2.6 User-Centric Privacy Solutions

*Future Direction:*

Further exploration into user-centric privacy solutions is crucial. This includes the development of decentralized identity platforms, self-sovereign identity systems, and user-friendly interfaces that empower individuals to have greater control over their data in IoT-based cloud systems.

*Rationale:*

Empowering users with control over their data aligns with the principles of privacy by design. As users become more informed and actively involved in privacy decisions, the overall effectiveness of privacy-preserving measures can be significantly enhanced.

The landscape of privacy preservation for IoT-based cloud systems is dynamic and presents both challenges and opportunities. Addressing the current gaps requires a multi-faceted approach that combines technical innovation with user empowerment and regulatory frameworks. As the IoT ecosystem continues to evolve, future research must remain adaptive, responsive, and collaborative to ensure that privacy-preserving measures

## 9. Conclusion

In conclusion, this survey paper has comprehensively addressed the intricate intersection of privacy preservation, artificial intelligence (AI), and IoT-based cloud systems. By elucidating the architectural foundations of IoT, the potential privacy threats, and a detailed examination of privacy-preserving techniques, the paper offers a holistic understanding of the challenges and solutions in securing sensitive data. The comparative analysis of privacy-preserving techniques and the exploration of case studies involving AI applications underscore the transformative impact of intelligent technologies in ensuring compliance, proactively managing risks, and fortifying the privacy landscape. Looking forward, future research avenues include the development of hybrid privacy-preserving models, advancements in explainable AI and federated learning, exploration of edge-driven privacy preservation, and the establishment of privacy-preserving regulations specific to IoT-based cloud systems. The ongoing evolution of smart AI-powered privacy policies and predictive privacy maintenance signifies a dynamic frontier in which technology continues to play a crucial role in safeguarding user data. As we advance, ethical considerations and user-centric solutions should remain at the forefront, ensuring that the integration of AI into privacy preservation aligns with societal values, transparency, and user empowerment. The collaborative efforts of researchers, practitioners, and regulatory bodies will be instrumental in shaping a future where privacy is not merely a compliance requirement but a foundational principle embedded in the fabric of connected systems.

**Acknowledgements**

This research was not funded by any grant.

**Author contributions**

Dhinakaran D: Conceptualization, Methodology, Writing-Original draft preparation, Udhaya Sankar S. M: Data curation, Software, Validation, Field study, Selvaraj D: Methodology, Writing-Reviewing and Editing, Edwin Raja S: Visualization, Investigation, Field study.

**Conflicts of interest**

The authors declare no conflicts of interest.